# HIGH PERFORMANCE SELECTABLE-VALUE TRANSPORTABLE HIGH DC VOLTAGE STANDARD

**Flavio Galliana[1), Roberto Cerri[2), Luca Roncaglione Tet[3)

1) *National Institute of Metrological Research, strada delle Cacce 91, 10135 TURIN – Italy* ( *f.galliana@inrim.it*)
2) *National Institute of Metrological Research, strada delle Cacce 91, 10135 TURIN – Italy* ( *r.cerri@inrim.it*)
3) *National Institute of Metrological Research, strada delle Cacce 91, 10135 TURIN – Italy* ( *l.roncaglione@inrim.it*)

**Abstract**

At National Institute of Metrological Research (INRIM), a selectable-value Transportable High dcVoltage Standard (THVS) operating in the range from 10 V to 100 V in steps of 10 V, was developed. This Standard was built to cover the lack of high level dc Voltage Standards at voltages higher than 10 V to employ as laboratory (local) or travelling Standards for Inter-Laboratory Comparisons (ILCs). A ground-mobile electronic technique was used to enhance the accuracy of the THVS at the higher values. The THVS shows better noise, better short-mid-term stability than top level dc Voltage and multifunction calibrators (MFCs) and better suitability and insensibility to be transported than these instruments. The project is extensible to 1000 V.

## 1. Introduction

Nowadays, the dc Voltage Standard is reproduced from the Standard of Time by means of the Josephson effect [1] while its maintenance is granted by Zener-diode-based dc Voltage Standards whose values are periodically updated repeating the Josephson effect [1–3]. These are excellent travelling Standards for their satisfactory rejection to physical shocks and temperature changes and their possibility to operate in battery mode. For this reason they are involved in interlaboratory comparisons (ILCs) at international and national level [4–7]. They are also used for the "artifact calibration" with which some high accuracy multifunction instruments can be calibrated and adjusted [8–12]. High accuracy dc Voltage Standards at values higher than 10 V are instead not available. Modern electrical calibration laboratories are equipped with dc Voltage calibrators and MFCs as reference Standards for dc Voltages up to 1000 V. These instruments assure satisfactory stability and accuracy, remote control and commercial availability. On the other hand, they can suffer of noise problems at their input stage [13]. When calibrators are involved in measurement setups with other sensitive instruments, common mode and power supply noises may cause measurement errors in particular at higher voltages due to their many internal circuits. In addition, the calibrators can be damaged during transports due to their dimensions and sensitivity to mechanical stresses. To overcome these problems, at National Institute of Metrological Research (INRIM) a modular multi-value high accuracy Transportable High dc Voltage Standard (THVS) operating from 10 V to 100 V in steps of 10 V was developed. It operates connected to the mains or in floating mode connected only to battery reducing noises. In fact, all its circuits can be supplied by means of a set of lead batteries that can be recharged when the device is not involved in measurements. The THVS was built with a ground mobile electronic technique aimed to enhance the accuracy of the THVS at the higher voltages. This paper shows the main features of the THVS, its characterization results in comparison with commercial top class dc Voltage calibrators and MFCs, its calibration and use[1] uncertainties as local or transportable Standard.

---

[1] The use uncertainty is the effective uncertainty that a standard or instrument introduces in the time period between two its calibrations when it is used to calibrate other standards or instruments. It normally includes uncertainty components

## 2. Description of the THVS

A block scheme of the THVS is shown in Fig. 1. It has a commercial circuit acting as internal voltage reference (10 Vdc), with temperature coefficient (TCR) $<0.3\times10^{-8}$/°C and estimated stability of $\pm\,0.5\times10^{-6}$/year. A principle scheme of the THVS is shown in Fig.2 while its external front view is shown in Fig. 3.

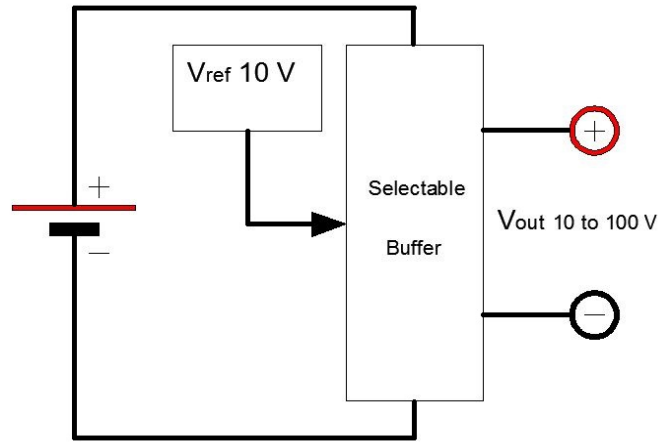

Fig. 1. Block scheme of the THVS.

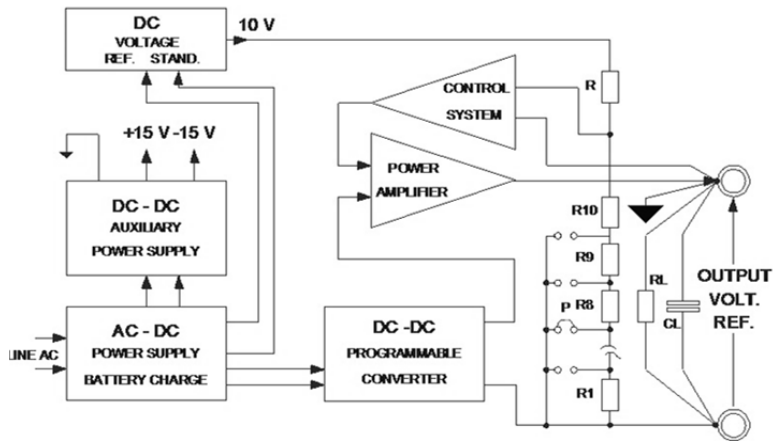

Fig. 2. THVS principle scheme.

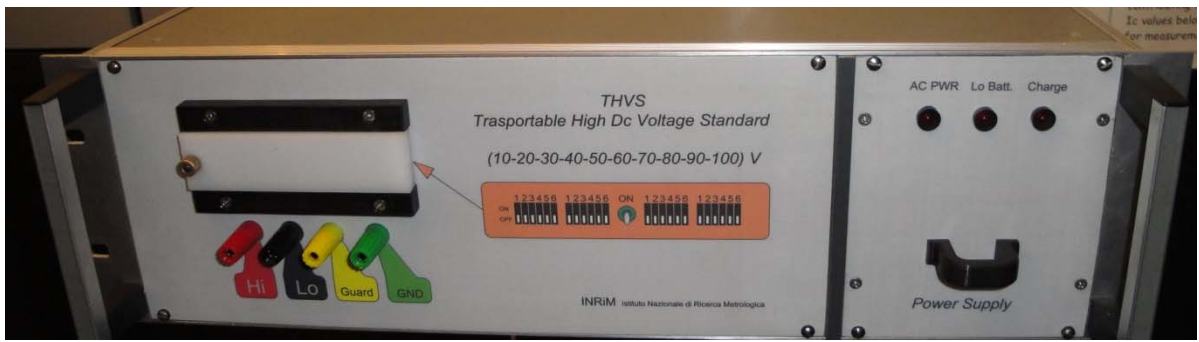

Fig. 3. Front view of the THVS.

---

due to its calibration, drift, environmental conditions and other influence parameters dependence. Similar concept was given in [14].

The THVS receives the correct supply voltages by means of precision low-ripple programmable dc-dc converter to obtain the selected output voltages. An additional low noise and high insulation dc-dc converter with outputs of ± 15 Vdc provides auxiliary voltages to control the output stage. This stage employs high voltage N-channel MOS components as power buffer. The control circuits are made with precision instrumentation amplifiers. The resistors of the internal divider are hermetically sealed high-precision Z–Foil with TCR lower than $0.4\times10^{-6}$/°C. The output stage is equipped with a protection system for maximum voltage and current. The technique to control the generated voltages is visible in Fig. 2 where the internal ground potential is driven to the high potential. This choice allows to increase the accuracy of the THVS at the higher voltages and to use components normally utilized for low voltages and available at lower cost. The electrical power required by the circuitry to control the electrical output of the THVS is supplied by three rechargeable sealed lead 6 V 5Ah batteries that ensure to the THVS an autonomous operation of some hours. Downstream the batteries, an accurate voltage regulator provides a dc voltage of 12 V assuring a correct supply of the dc-dc converter. Its characteristics are: noise voltage $\cong$ 30 µVpp, TCR $\cong 1\times10^{-5}$/°C and output current of 0.5 A.

*2.1. THVS components and characteristics*

The resistors involved in the THVS are Vishay VSRJ type 10 kΩ with tolerance of ± 0.05 %), TCR lower than $0.4\times10^{-6}$/°C, thermal electromotive force (EMF) of ± 0.05 µV/V, power at 70°C of $\cong$ 0.3 W. For the 10 V and 20 V were inserted two Vishay VH102Z type 10 kΩ with same tolerance and power but with TCR lower than $0.2\times10^{-6}$/°C to improve the stability of these two values of the THVS.

The main electronic component is a MOSFET VISHAY mod. IRFR220 with $V_{drain\text{-}Source(DS)}$) of 200 V, $R_{DS\,on}$ of 0.8 Ω and $I_{DSon}$ of 3A. This power stage was realized with technologies Hexfet (Power MOSFET) designed to operate at constant power regardless of the set output voltage. Other significant components are:

- A dc-dc switching with programmable output from 18 V to 110 Vdc with instability $< 25\times10^{-6}$/° C, peak to peak ripple at full load: <0.01%, frequency 80 kHz÷180 kHz and high insulation capacity;
- A control circuit for the power section made with ultraprecision operational amplifiers (offset voltage < 10µV, offset drift < 0.1 µV /°C).

*2.2. THVS specifications*

The specifications of the THVS are:
- Output voltages: from 10 V to 100 V with steps of 10 V selectable by means of a switch placed on the front panel;
- Output currents ≤ 5 mA;
- Output noise at 100 V ≤ 6.9 µV rms operating connected to the mains and ≤ 5.0 µV operating connected only to battery. These values have to be compared with 155 µV and 153 µV of a top class dc Voltage calibrator and a top class MFC as declared by the manufacturers;
- Evaluated 24h mean stability of $1.2\times10^{-8}$ at 100 V to compare with $3.8\times10^{-7}$/24h and $5.0\times10^{-8}$/24h of two dc Voltage calibrators and with $1.0\times10^{-7}$/24h and $2.5\times10^{-7}$/24h of two MFCs;
- Evaluated mid-term stability of $1.6\times10^{-6}$ at six months at 100 V.

*2.3. Thermal features of the THVS.*

To investigate the temperature behaviour of the THVS in the typical temperature range of electrical calibration laboratories (23 ± 1 °C), its temperature coefficient around 23 °C was evaluated soon after its assembly. The THVS was measured, after suitable stabilization, at (22, 23 and 24) °C in a settable temperature laboratory. The temperature coefficient resulted

≅ − 8.3×10$^{−8}$/°C. To further minimize the temperature dependence of the THVS, a thermal compensator maintaining the temperature inside the THVS at 37.70 °C was added. It rejects the temperature changes due to different load effects of the different voltages and due to the external temperature variations. By means of the compensator the temperature fluctuation in the THVS is maintained within ± 0.15 °C at the temperature conditions of an electrical laboratory lowering the corresponding uncertainty component. The action of this compensator allows a faster stabilization of a new selected voltage and, as the THVS is maintained always at this temperature, its humidity dependence is minimized.

### 3. Comparison with DC Voltage calibrators and MFCs

Two alternative tests were carried out to compare the THVS at 100 V with high accuracy dc Voltage calibrators and MFCs in their dc Voltage mode. In the first test, the THVS, a dc Voltage calibrator and a MFC were compared at 100 V connecting them alternatively to the same high accuracy DMM in its 100 V range. The DMM measurements at 100 V were computed nulling its 0 V readings. The measurements were made in a shielded laboratory thermo-regulated at (23 ± 0.3) °C and at a relative humidity of (40 ± 10) %. The 12h measurement results of the three standards are shown in Fig. 4. In this case the THVS was connected to the mains. The measurements relative spreads, evaluated as standard deviations, were 3.9×10$^{−8}$, 1.0×10$^{−7}$ and 8.3×10$^{−8}$ respectively for the THVS, for the dc Voltage calibrator and for the MFC. These values include the DMM contribution that was considered stable in the three evaluations as the better available DMM was selected. The lowest drift was obtained by the THVS. In Fig. 5 the 3h measurement behaviour of all the dc Voltages provided by the THVS in floating mode (battery-fed) are reported. This time was chosen as it is the typical time necessary to calibrate dc Voltage Standards in floating mode. The lowest relative spread (3.2×10$^{−8}$) was obtained at 100 V confirming the lower instability and noise of the THVS operating in battery mode.

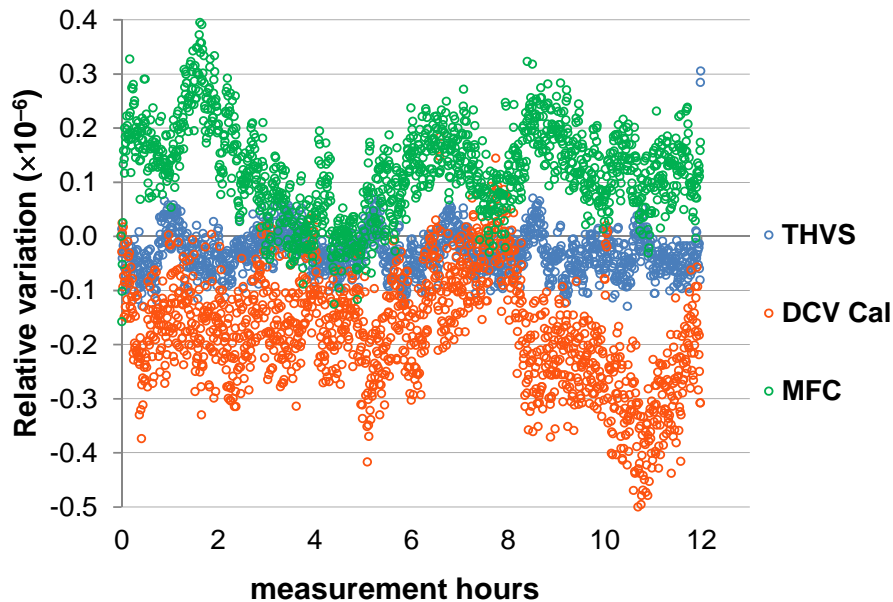

Fig. 4. 12h relative spread and drift comparison among the THVS, a dc Voltage calibrator and a MFC at 100 V reading with a high accuracy DMM.

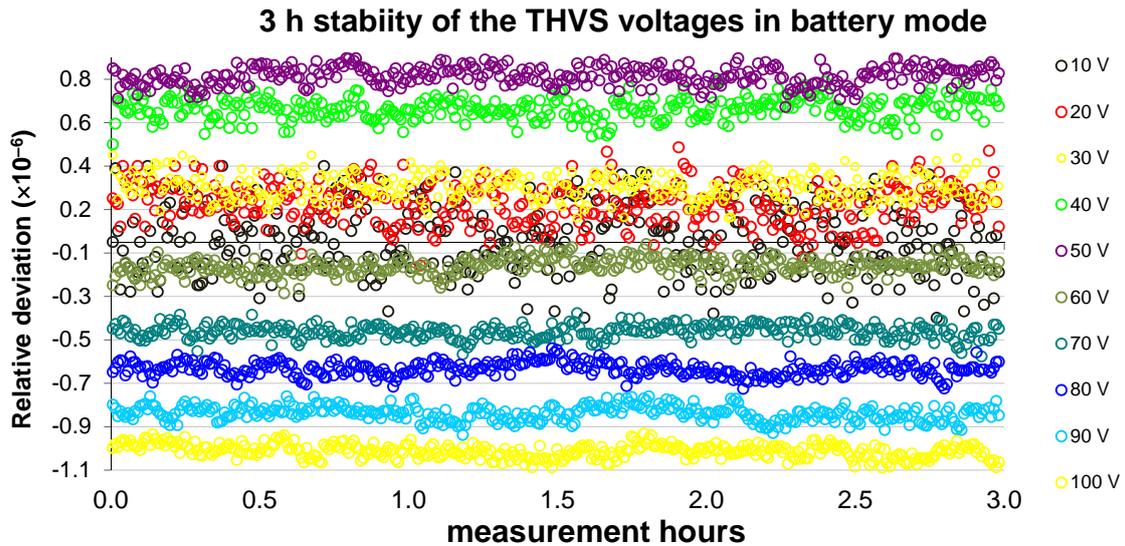

Fig. 5. 3h stability comparison among the dc Voltages of the THVS operating in battery mode.

With the second test, three DMMs with similar noise and intrinsic repeatability were selected. Then, the THVS was compared at 100 V in a 24h time-period after a same stabilization period before with two top class dc Voltage calibrators (Fig. 6) and after with two MFCs (Fig. 7). The comparison were made connecting the instruments at the same time to the three selected DMMs. The measurements were carried out during the weekend to avoid disturbs due to presence of the operators. In these measurements the compared instruments were subjected to the same environmental fluctuations. The THVS was connected to the mains. In the first case the relative spreads were $3.6 \times 10^{-8}$, $1.2 \times 10^{-7}$ and $6.8 \times 10^{-8}$ respectively for the THVS and for the two dc Voltage calibrators. The lowest drift was obtained by the THVS with a relative deviation of $1.1 \times 10^{-8}/24h$ vs. $3.8 \times 10^{-7}/24h$ and $5.0 \times 10^{-8}/24h$ for the two dc Voltage calibrators.

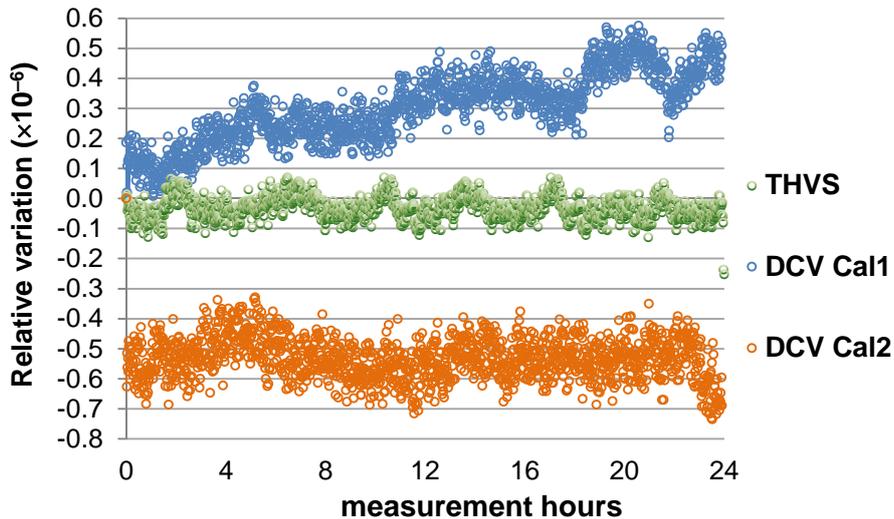

Fig. 6. 24h relative spread and drift of the THVS and of two dc Voltage calibrators at 100 V with the three DMMs test.

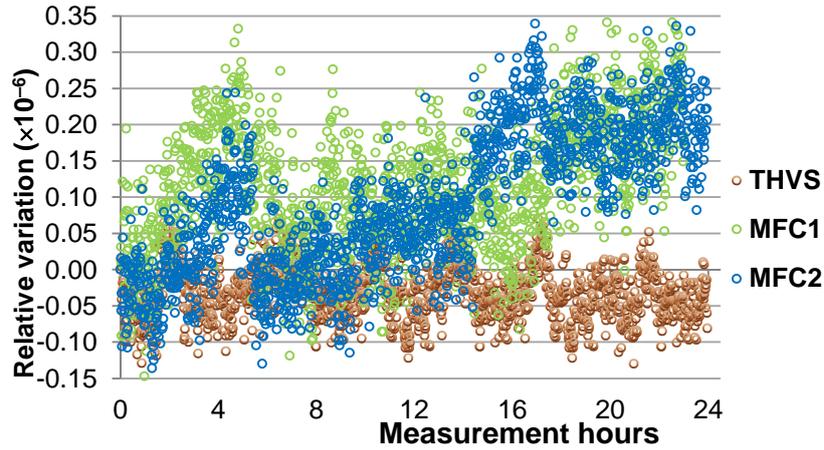

Fig. 7. 24h relative spread and drift of the THVS and of two MFCs at 100 V with the three DMMs test.

In the second case the relative spreads were $3.7 \times 10^{-8}$, $2.3 \times 10^{-7}$ and $1.0 \times 10^{-7}$ respectively for the THVS and for the two MFCs. The lowest drift was obtained by the THVS with a relative drift of $1.2 \times 10^{-8}$/24h vs. $1.0 \times 10^{-7}$/24h and $2.5 \times 10^{-7}$/24h for the two MFCs.

### 4. Calibration of the THVS

The THVS is calibrated with an opposition method by means of the measurement setup shown in Fig. 8. The THVS has to be connected to the input of a high accuracy INRIM dc Voltage divider [15] set in suitable ratio. Then the divider output voltage is compared with a INRIM Zener dc Voltage Standard (10 V) calibrated vs. the INRIM national dc Voltage Standard. Between the divider output and the Voltage Standard, a high precision DMM measures the unbalance voltages.

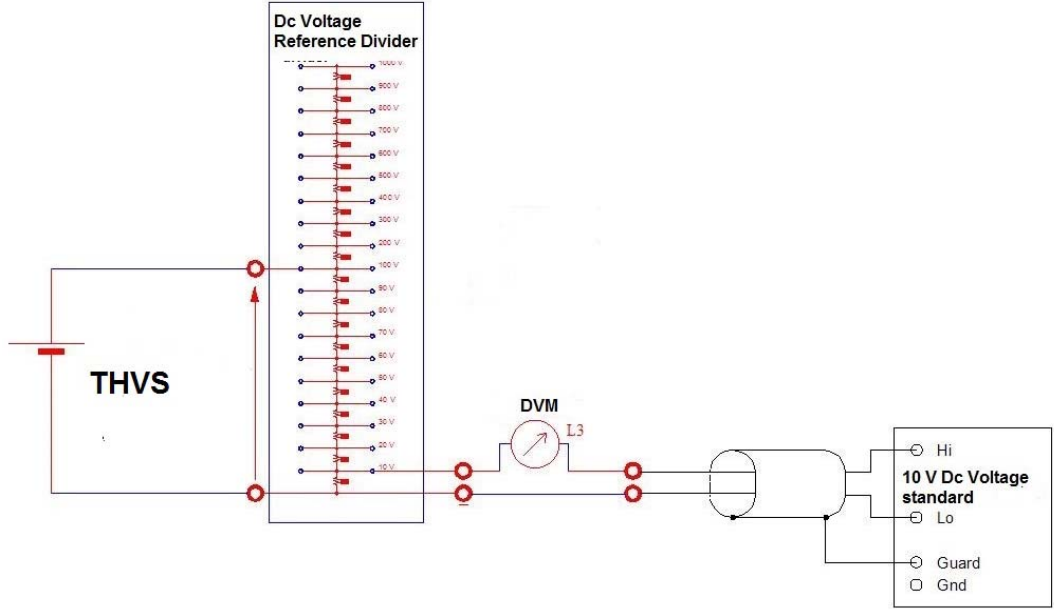

Fig. 8. Measurement setup to calibrate the THVS.

Thus, the THVS value is:

$$V_{THVS} = \frac{V_s + v}{D} \quad (1)$$

where $V_s$ is the value of the 10 V dc Voltage Standard, $v$ the voltage unbalance and $D$ the divider dc Voltage ratio. In Fig. 9 a photo of the measurement setup of fig. 8 is shown.

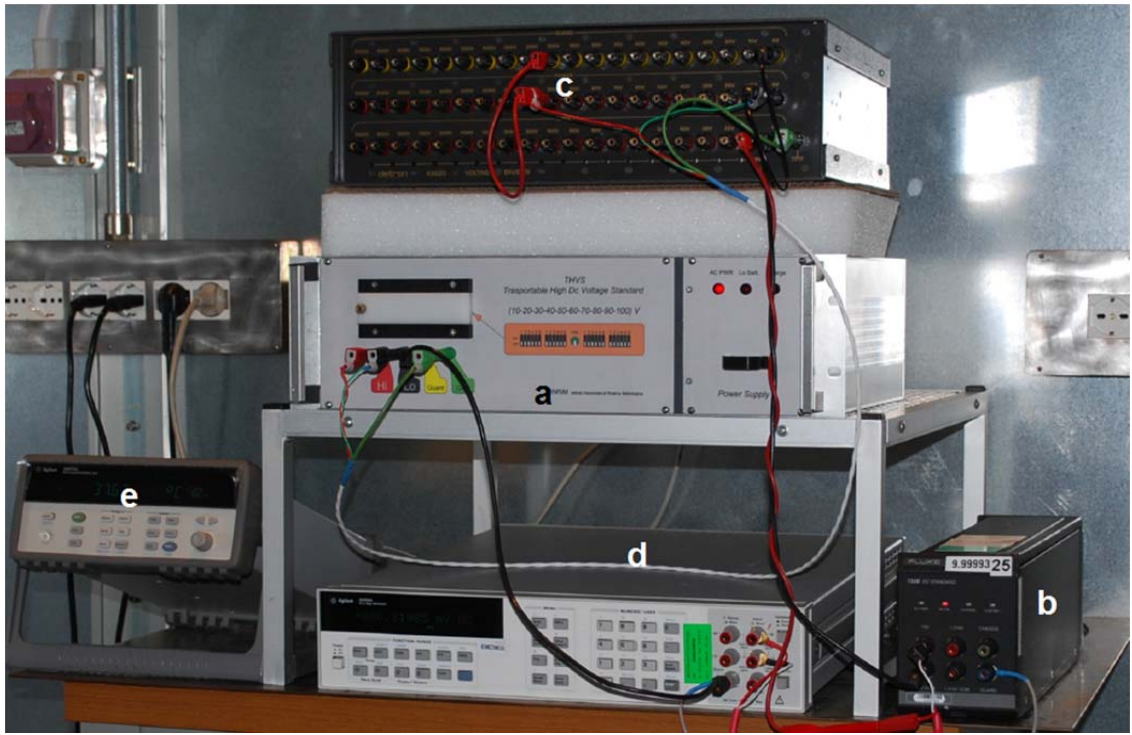

Fig. 9. View of the measurement setup to calibrate the THVS. a) THVS, b) 10 V reference Standard. c) dc Voltage standard divider, d) DMM used to measure the unbalance voltages, e) temperature-meter to acquire the temperature inside the THVS.

## 5. Evaluation on the uncertainties of the THVS

### 5.1. THVS calibration uncertainty

According to the paragraph 4 and to (1) in Table 1 a relative uncertainty budget for the calibration of the THVS at 100 V is given.

Table 1. THVS calibration relative uncertainty budget at 100 V.

| Source | type | $1\sigma$ ($\times 10^{-7}$) |
|---|---|---|
| Reference 10 V calibration | B | 2.5 |
| Reference drift | B | 0.6 |
| Ref temp. dependence | B | 0.2 |
| Ref humidity dependence | B | negl. |
| DMM accuracy | B | 0.2 |
| DMM calibration | B | 0.1 |
| Unbalance repeatability | A | 0.9 |
| Divider calibration | B | 1.0 |
| Divider drift | B | 0.1 |
| Divider temp. dependence | B | 0.1 |
| Total RSS | | 2.9 |

For a 95 % confidence level the calibration relative uncertainty of the THVS at 100 V is then $5.9 \times 10^{-7}$. According to these uncertainty components, in Table 2 the expanded calibration relative uncertainties of the THVS are listed.

Table 2. THVS calibration relative expanded uncertainty for each voltage.

| Voltage (V) | Expanded uncertainty (×10⁻⁷) |
|---|---|
| 10 | 7.0 |
| 20 | 7.1 |
| 30 | 7.0 |
| 40 | 6.4 |
| 50 | 6.3 |
| 60 | 6.1 |
| 70 | 6.1 |
| 80 | 6.0 |
| 90 | 5.9 |
| 100 | 5.9 |

*5.2. Mid-term stability of the THVS*

The THVS was also measured about every week at various voltages with the measurement setup of Fig. 8 to evaluate its mid-term stability. The THVS showed a smooth mean linear decreasing drift of $1.6\times10^{-6}$ in six months at 100 V (Fig. 10). This drift can be presumably due to the incomplete stabilization of the THVS internal components. The measurements will continue to establish the long-term regimen drift of the THVS.

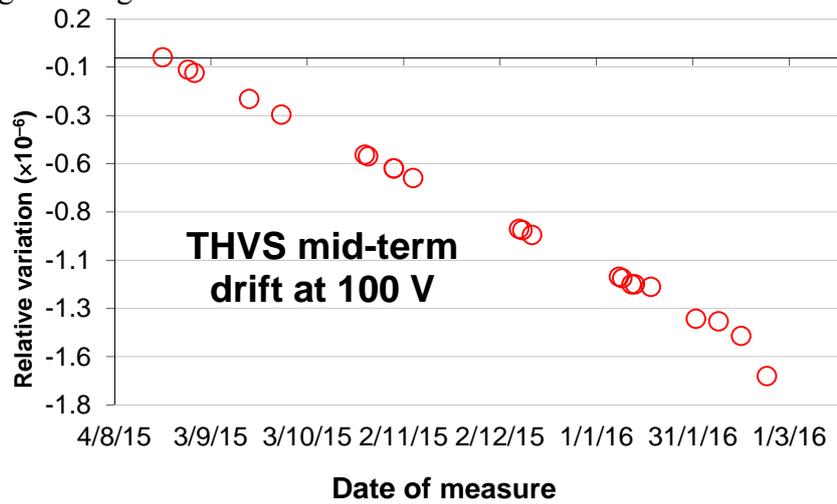

Fig. 10. drift of the THVS at 100 V since its assembly.

*5.3. Use uncertainty of the THVS*

in Table 3 a relative use uncertainty budget of the THVS at 100 V is reported. It was assumed to use the THVS as dc Voltage Standard for 180 days without recalibration.

Table 3. THVS mid-term use relative uncertainty budget at 100 V.

| Source | type | 1 σ (×10⁻⁶) |
|---|---|---|
| calibration | B | 0.29 |
| drift | B | 0.46[2] |
| Temp/hum dependence | B | 0.0 |
| noise | B | 0.1 |
| Total RSS | | 0.55 |

---

[2] This uncertainty component was evaluated according to [16], clause F.2.2.2.

For a 95% confidence level the use relative uncertainty of the THVS at 100 V is then about $1.1\times10^{-6}$. According to these uncertainty components, in Table 4 the expanded relative use uncertainties of the THVS are listed. These uncertainty values are valid using the THVS as laboratory Standard for at least six months after calibration.

Table 4. THVS use relative expanded uncertainties as laboratory Standard for each voltage.

| Voltage (V) | Expanded uncertainty ($\times 10^{-6}$) |
|---|---|
| 10 | 1.9 |
| 20 | 1.8 |
| 30 | 1.7 |
| 40 | 1.5 |
| 50 | 1.4 |
| 60 | 1.3 |
| 70 | 1.3 |
| 80 | 1.1 |
| 90 | 1.1 |
| 100 | 1.1 |

### 5.4. Transport effect on the THVS

In a hypothetical ILC the THVS could be transported by car, van, or plane and maintained for several hours or some days in not controlled environment conditions till to the arrival to the participating laboratories. In our test the THVS was measured at INRIM before the transport; after it was transported in a suitable package by car with 2-3h of travel, successively maintained in uncontrolled temperature condition for at least 24h. Then it was placed in a thermo-regulated laboratory for 24h before to be measured again. The observed maximum relative difference between the measurements made before and after the transport, analysing all voltages, was $2.0\times10^{-7}$. In Table 5 the use relative uncertainties as travelling Standard, obtained adding to the use ones as laboratory Standard the transport uncertainty component, are listed.

Table 5. THVS relative expanded use uncertainties as travelling Standard for each voltage.

| Voltage (V) | Expanded uncertainty ($\times 10^{-6}$) |
|---|---|
| 10 | 1.9 |
| 20 | 1.8 |
| 30 | 1.7 |
| 40 | 1.6 |
| 50 | 1.5 |
| 60 | 1.4 |
| 70 | 1.4 |
| 80 | 1.2 |
| 90 | 1.2 |
| 100 | 1.1 |

### 6. Conclusions

As it can be seen in Tables 2, 4 and 5 the THVS shows better uncertainties at higher voltages due to its construction technique. In the characterization and stability tests it showed lower noise, better short and mid-time stability and measurement repeatability than top class dc Voltage calibrators and MFCs. This is a significant result as the THVS is not completely actively thermo-regulated. The THVS is suitable to act as top level dc Voltage laboratory Standard or travelling multiple-value Standard for national ILCs. Future aim will be the evaluation of the THVS pressure dependence for its possible involvement also in international ILCs. In a few months the THVS will be implied in an ILC with some high level accredited italian laboratories. This ILC will be the occasion to further verify its stability, transport effect and environmental dependence. The project of the THVS is extensible in future adding to the actual realization a module with a Voltage source providing a dc selectable Voltage ranging from 100 V to 1000 V with steps of 100 V and with a complete active thermo-regulation.

**Acknowledgements**

The authors wish to thank F. Francone for his help in the construction of mechanical details of the THVS.